\documentclass[preprint,amsmath,amssymb,aps,showkeys]{revtex4}
\usepackage[english]{babel}
\usepackage{graphicx}
\usepackage{graphics}
\usepackage{amsmath}
\usepackage{dcolumn}
\usepackage{amssymb}
\usepackage{bm}

\begin{document}
\title{Revisiting the missing He-McKellar-Wilkens geometric quantum phase}
\author{K. Bakke}
\email[E-mail address: ]{kbakke@fisica.ufpb.br}
\homepage[Orcid: ]{https://orcid.org/0000-0002-9038-863X}
\affiliation{Departamento de F\'isica, Universidade Federal da Para\'iba, Caixa Postal 5008, 58051-900, Jo\~ao Pessoa, PB, Brazil.}

\begin{abstract}

We discuss the missing He-McKellar-Wilkens geometric quantum phase in Landau levels for a neutral particle with a permanent electric dipole moment in the presence of an infinity wall. We also discuss the influence of the missing He-McKellar-Wilkens geometric quantum phase on the energy levels of a neutral particle confined to a one-dimensional quantum ring. Further, we explore this influence of the missing He-McKellar-Wilkens geometric quantum phase on the energy levels by calculating the persistent spin currents.

\end{abstract}


\maketitle

\section{Introduction}

Although magnetic monopoles are not elementary particles, they draw attention to the beauty of the symmetry between the magnetic and electric fields in the Maxwell equations. Dirac \cite{dirac} proposed the existence of magnetic monopoles with the purpose of achieving this symmetry in the Maxwell equations. The hypothesis of magnetic monopoles can be explored from the theoretical point of view through the duality transformations in the Maxwell equations \cite{abdual,furdual,griff2,ex1}. It has also been used in the search for unified gauge theories \cite{magcharge7,magcharge8,magcharge9,magcharge10,magcharge11}. In view of experiments, we can cite the experiments made in Refs. \cite{magexp1,magexp2,magexp3,magexp4,magexp5,magexp6,magexp,magexp7}. Another perspective is given in condensed matter physics with the aim of finding analogues of the magnetic monopoles. For instance, magnetic monopole analogues have been found in spin ice systems \cite{magcharge2,magcharge3,ex2,ex3}, synthetic magnetic field \cite{diracmono}, skyrmion lattice \cite{diracmono4}, defects in liquid crystals \cite{diracmono2} and crystal momentum space of solids \cite{diracmono3}.

Studies of geometric quantum phases have also been inspired by the Dirac monopoles. The dual effect of the Aharonov-Bohm effect \cite{abdual,furdual} and the scalar Aharonov-Bohm effect for a neutral particle with a magnetic quadrupole moment \cite{fb7} are examples of them. The most well known study of geometric quantum phases in this theme deals with a neutral particle with a permanent electric dipole moment that interacts with a magnetic field produced by a linear distribution of magnetic charges. This geometric quantum phase is known as the He-McKellar-Wilkens effect \cite{hmw,hmw2}. By considering a nonuniform magnetization inside a long ferromagnetic wire, Tkachuk \cite{tka} proposed a way of reproducing the radial magnetic field that gives rise to the He-McKellar-Wilkens effect \cite{hmw,hmw2}. Later, Furtado and de Lima Ribeiro \cite{anan1c} explored Tkachuk's magnetic field configuration and studied geometric quantum phases for neutral particles.

With the focus on bound states, Ribeiro {\it et al} \cite{lin} proposed a radial magnetic field produced by a uniform magnetic charge density with the purpose of achieving the Landau quantization \cite{landau} for a neutral particle with a permanent electric dipole moment. Inspired by Ribeiro {\it et al} proposal \cite{lin}, bound states for neutral particles have been studied based on the interaction of magnetic quadrupole moment with radial magnetic fields produced by nonuniform magnetic charge densities, and axial electric fields produced by magnetic current densities \cite{vb,vb2}. Recently, by considering an infinitely long non-conducting cylinder with a cylindrical cavity and a uniform distribution of the magnetic charges inside it, a geometric quantum phase for a neutral particle has been proposed associated with the missing magnetic charge per unit length inside the cylindrical cavity \cite{b}. This geometric quantum phase can be viewed as the inverse of He-McKellar-Wilkens proposal \cite{hmw,hmw2} and yields an Aharonov-Bohm-type effect for bound states \cite{ab,pesk}. We have called the missing He-McKellar-Wilkens geometric phase \cite{b}.

In this work, we revisit the interaction of a radial magnetic field produced by a uniform distribution of the magnetic charges inside the long non-conducting cylinder with a cylindrical cavity with the permanent electric dipole moment of the neutral particle. We bring two different perspectives on studying the missing He-McKellar-Wilkens geometric phase \cite{b}. In the first one, we show that the missing He-McKellar-Wilkens geometric phase can appear in the Landau-He-McKellar-Wilkens quantization \cite{lin} even though an infinity wall is located at the inner radius of the non-conducting cylinder. Later, we study the missing He-McKellar-Wilkens geometric phase when the neutral particle is confined to a one-dimensional quantum ring. Furthermore, we study the persistent spin currents \cite{perspin,perspin2,perspin3,perspin4} associated with the missing He-McKellar-Wilkens geometric phase.

The structure of this paper is: in section II, we start with a brief review of the missing He-McKellar-Wilkens geometric phase \cite{b}. Then, we show that an Aharonov-Bohm-type effect for bound states can be achieved in the Landau-type system proposed in Ref. \cite{lin} associated with the missing He-McKellar-Wilkens geometric phase. In addition, we obtain the persistent spin currents with regard to the missing He-McKellar-Wilkens geometric phase; in section III, we study the confinement of a neutral particle to a one-dimensional quantum ring. Thus, we discuss the Aharonov-Bohm-type effect for bound states and the persistent spin currents associated with the missing He-McKellar-Wilkens geometric phase; in section IV, we present our conclusions.

\section{Aharonov-Bohm-type effect in a Landau-type system}

The interaction of a neutral particle with a permanent electric dipole moment with electric and magnetic fields is described by the Schr\"odinger-Pauli equation \cite{anan1a,ana,hmw,hmw2,b} (with units: $\hbar=1$ and $c=1$):
\begin{eqnarray}
\mathcal{E}\psi=\frac{1}{2m}\left[\hat{\pi}-d\,\vec{\sigma}\times\vec{B}\right]^{2}\psi-\frac{d^{2}\,B^{2}}{2m}\,\psi-\frac{d}{2m}\left(\vec{\nabla}\cdot\vec{B}\right)\psi+d\,\vec{\sigma}\cdot\vec{E}\,\psi,
\label{1.1}
\end{eqnarray}
where $d$ is the permanent electric dipole moment of the neutral particle, $\vec{E}$ and $\vec{B}$ are the electric and magnetic fields, respectively. Besides, $\vec{\sigma}$ corresponds to the Pauli matrices, which satisfy the relation $\left(\sigma^{i}\,\sigma^{j}+\sigma^{j}\,\sigma^{i}\right)=2\,\delta^{ij}$. We also have the presence of the the operator $\hat{\pi}$, whose components are given by $\hat{\pi}_{k}=-\frac{i}{h_{k}}\,\partial_{k}-\frac{1}{2\,r}\,\sigma^{3}\,\delta_{\varphi\,k}$ \cite{bf15}. Observe that the parameter $h_{k}$ is known as the scale factor of this coordinate system. With regard to the cylindrical symmetry, the scale factors are $h_{r}=h_{1}=1$, $h_{\varphi}=h_{2}=r$ and $h_{z}=h_{3}=1$ \cite{arf}.

Recently, we have proposed the study of a geometric quantum phase which yields an analogue of the He-McKellar-Wilkens geometric quantum phase \cite{hmw,hmw2}. This geometric quantum phase has stemmed from the absence of magnetic magnetic charges per unit length inside the cylindrical cavity of an infinitely long non-conducting cylinder. We have called the missing He-McKellar-Wilkens geometric quantum phase \cite{b}. We have achieved this geometric quantum phase by considering the magnetic field produced by a uniform distribution of the magnetic charges inside an infinitely long non-conducting cylinder, where the cylinder has an inner radius $r_{0}$. Hence, the magnetic field in the region $r\,\geq\,r_{0}$ is $\vec{B}=\left[\frac{\rho_{m}}{2}\,r-\frac{\rho_{m}\,r_{0}^{2}}{2\,r}\right]\,\hat{r}$, where $\rho_{m}$ is the uniform volume magnetic charge density and $\hat{r}$ is the unit vector in the radial direction. From Eq. (\ref{1.1}), we have an effective vector potential defined by  $\vec{A}_{\mathrm{HMW}}=d\,\vec{\sigma}\times\vec{B}$ \cite{hmw,hmw2}. In this way, by assuming that the electric dipole moment of the neutral particle is aligned along the $z$-axis, the effective vector potential becomes $\vec{A}_{\mathrm{HMW}}=\vec{A}_{\mathrm{HMW}1}-\vec{A}_{\mathrm{HMW}2}$, where
\begin{eqnarray}
\vec{A}_{\mathrm{HMW}1}=\frac{d\,\rho_{m}\,r}{2}\,\sigma^{3}\,\hat{\varphi};\,\,\,\,\,\vec{A}_{\mathrm{HMW}2}=\frac{d\,\rho_{m}\,r_{0}^{2}}{2\,r}\,\sigma^{3}\,\hat{\varphi},
\label{1.3}
\end{eqnarray}
where $\hat{\varphi}$ is a unit vector in the azimuthal direction. From the contribution given by $\vec{A}_{\mathrm{HMW}2}$, we have obtained the missing He-McKellar-Wilkens geometric quantum phase \cite{b}: 
\begin{eqnarray}
\Phi_{\mathrm{MHMW}}=\oint\vec{A}_{\mathrm{HMW}2}\cdot d\vec{r}=\pi\,d\,\rho_{m}\,r_{0}^{2}\,\sigma^{3}.
\label{1.4}
\end{eqnarray}

On the other hand, the contribution given by $\vec{A}_{\mathrm{HMW}1}$ yields an effective uniform magnetic field $\vec{B}_{\mathrm{eff}}=\vec{\nabla}\times\vec{A}_{\mathrm{HMW}1}=d\,\rho_{m}\,\sigma^{3}\hat{z}$ in the region $r\geq r_{0}$ analogous to that of Ref. \cite{lin}. In this sense, the neutral particle is in a Landau-type system \cite{lin}. Furthermore, as we have shown in Ref. \cite{b}, the solution to the Schr\"odinger-Pauli equation (\ref{1.1}) in the region $r\,\geq\,r_{0}$ is given by $\psi\left(r,\,\varphi,\,z\right)=e^{i\left(\ell+1/2\right)\varphi}e^{ip_{z}\,z}\,f\left(r\right)$. Note that $\ell=0,\pm1,\pm2,\ldots$ is the eigenvalue of $\hat{L}_{z}$ and $p_{z}$ is the eigenvalue of $\hat{p}_{z}$. Thus, the radial equation is 
\begin{eqnarray}
f''+\frac{1}{r}\,f'-\frac{\gamma^{2}}{r^{2}}\,f-\frac{m^{2}\varpi^{2}}{4}\,r^{2}\,f+\tau\,f=0.
\label{1.5}
\end{eqnarray} 
The parameters $\varpi$, $\gamma$ and $\tau$ are defined as follows \cite{b}:
\begin{eqnarray}
\varpi&=&\frac{d\,\rho_{m}}{m};\nonumber\\
\gamma&=&\ell+\frac{1}{2}\left(1-s\right)-\frac{\Phi_{\mathrm{MHMW}}}{2\pi};\label{1.6}\\
\tau&=&2m\mathcal{E}-s\,m\varpi\,\gamma+m\,\varpi,\nonumber
\end{eqnarray}
where we have taken $p_{z}=0$. By defining $y=\frac{m\,\varpi}{2}\,r^{2}$, Eq. (\ref{1.5}) becomes:
\begin{eqnarray}
y\,f''+f'-\frac{\gamma^{2}}{4y}\,f-\frac{y}{4}\,f+\frac{\tau}{2m\varpi}\,f=0.
\label{1.7}
\end{eqnarray}

Let us work with the solution to Eq. (\ref{1.7}) regular at $y\rightarrow\infty$. Therefore, it is given in terms of the confluent hypergeometric function of the second kind $U\left(a,\,b;\,y\right)$, i.e,
\begin{eqnarray}
f\left(y\right)=e^{-y/2}\times y^{\left|\gamma\right|/2}\times U\left(\frac{\left|\gamma\right|}{2}+\frac{1}{2}-\frac{\tau}{2m\varpi},\,\left|\gamma\right|+1;\,y\right).
\label{1.8}
\end{eqnarray}

In order to study the interaction of the permanent electric dipole moment of the neutral particle with the magnetic field $\vec{B}=\left[\frac{\rho_{m}}{2}\,r-\frac{\rho_{m}\,r_{0}^{2}}{2\,r}\right]\,\hat{r}$ in the region $r\,\geq\,r_{0}$, we have assumed that the existence of an infinity wall at $r=r_{0}$ \cite{b}. Therefore, we have the following boundary condition:
\begin{eqnarray}
f\left(y_{0}\right)=0,
\label{1.9}
\end{eqnarray}
where $y_{0}=\frac{m\,\varpi\,r^{2}_{0}}{2}$ is defined for $r=r_{0}$.

In this work, we explore the boundary condition from a different point of view. In contrast to Ref. \cite{b}, our focus is on the case where $y_{0}\ll1$. We wish to deal with the function $U\left(a,\,b;\,y_{0}\right)$ for the case where $y_{0}\ll1$ and $b>1$. In this case, the function $U\left(a,\,b;\,y_{0}\right)$ can be written in the form \cite{abra}:
\begin{eqnarray}
U\left(a,\,b;\,y_{0}\right)&\propto&\frac{\Gamma\left(b-a\right)}{\Gamma\left(a\right)}\,y_{0}^{1-b},
\label{1.10}
\end{eqnarray}
where $\Gamma\left(a\right)$ is the Gamma function. Note that $a=\frac{\left|\gamma\right|}{2}+\frac{1}{2}-\frac{\tau}{2m\varpi}$, while $b=\left|\gamma\right|+1$ ($b\,>\,1$). Hence, by substituting Eqs. (\ref{1.10}) and (\ref{1.8}) into the boundary condition (\ref{1.9}), we have that the boundary condition is satisfied only if $\Gamma\left(a\right)\rightarrow\infty$. This occurs when $a=-n$, where $n=0,1,2,3,\ldots$. Thereby, we obtain the energy levels:
\begin{eqnarray}
\mathcal{E}_{n,\,\ell}&=&\varpi\left[n+\frac{\left|\gamma\right|}{2}+s\,\frac{\gamma}{2}\right]\nonumber\\
\nonumber\\
&=&\varpi\left[n+\frac{1}{2}\left|\ell+\frac{1}{2}\left(1-s\right)-\frac{\Phi_{\mathrm{MHMW}}}{2\pi}\right|+\frac{s}{2}\left(\ell+\frac{1}{2}\left(1-s\right)-\frac{\Phi_{\mathrm{MHMW}}}{2\pi}\right)\right].
\label{1.11}
\end{eqnarray}
where $n=0,1,2,3,\ldots$ is called the radial quantum number.

Hence, the energy levels (\ref{1.11}) also stem from the interaction of the permanent electric dipole moment of the neutral particle with the inhomogeneous magnetic field $\vec{B}=\left[\frac{\rho_{m}}{2}\,r-\frac{\rho_{m}\,r_{0}^{2}}{2\,r}\right]\,\hat{r}$ in the region $r\geq r_{0}$. In contrast to those obtained in Ref. \cite{b}, when $y_{0}\ll1$, the energy levels are analogous to the Landau-He-McKellar-Wilkens levels \cite{lin}, and there is no upper limit to the radial quantum number. However, the energy levels (\ref{1.11}) are influenced by the missing He-McKellar-Wilkens geometric phase $\Phi_{\mathrm{MHMW}}$, which does not occur with the Landau-He-McKellar-Wilkens levels \cite{lin}. Besides, the missing He-McKellar-Wilkens geometric phase breaks the degeneracy of the Landau-He-McKellar-Wilkens levels \cite{lin}. Therefore, the absence of magnetic flux inside the region $r\,<\,r_{0}$ or the missing He-McKellar-Wilkens geometric phase $\Phi_{\mathrm{MHMW}}$ influences the energy spectrum, and thus, it yields an Aharonov-Bohm-type effect for bound states \cite{ab,pesk}.

Furthermore, the energy levels (\ref{1.11}) are a periodic function of the missing He-McKellar-Wilkens geometric phase $\Phi_{\mathrm{MHMW}}$. As an example, when the periodicity is $\phi_{0}=2\pi$ and $\ell>0$, the quantum states determined by $\left\{n,\,\ell,\,s=+1\right\}$ and $\left\{n,\,\ell-1,\,s=+1\right\}$ are degenerate because $\mathcal{E}_{n,\,\ell}\left(\Phi_{\mathrm{MHMW}}+2\pi\right)=\mathcal{E}_{n,\,\ell-1}\left(\Phi_{\mathrm{MHMW}}\right)$. Another example is for the periodicity $\phi_{0}=-2\pi$ and $\ell>0$. Thus, the quantum states determined by $\left\{n,\,\ell,\,s=+1\right\}$ and $\left\{n,\,\ell+1,\,s=+1\right\}$ are degenerate, i.e., $\mathcal{E}_{n,\,\ell}\left(\Phi_{\mathrm{MHMW}}-2\pi\right)=\mathcal{E}_{n,\,\ell+1}\left(\Phi_{\mathrm{MHMW}}\right)$.

\subsection{Persistent spin currents}

Due to the fact that the energy levels (\ref{1.11}) are a periodic function of the geometric quantum phase $\Phi_{\mathrm{MHMW}}$, persistent currents can appear in the system \cite{by}. According to Refs. \cite{perspin,perspin2,perspin3,perspin4}, the persistent currents that appear in a neutral particle system associated with the Aharonov-Casher geometric phase \cite{ac} are called persistent spin currents. Other studies of persistent spin currents associated with the Aharonov-Casher geometric \cite{ac} have been made in two-dimensional quantum ring \cite{bf18} and in Dirac oscillator \cite{bf3}. Since we are dealing with a spin-$1/2$ neutral particle, the dependence of the energy levels (\ref{1.11}) on the missing He-McKellar-Wilkens geometric quantum phase $\Phi_{\mathrm{MHMW}}$ can yield persistent spin currents. By assuming that the temperature is $T=0$, the persistent spin currents are obtained from the Byers-Yang relation \cite{by,dantas,bf3,bf18}:
\begin{eqnarray}
\mathcal{I}=-\sum_{n,\,\ell}\frac{\partial \mathcal{E}_{n,\,\ell}}{\partial\Phi_{\mathrm{MHMW}}}.
\label{3.1}
\end{eqnarray}

From the energy levels (\ref{1.11}), the persistent spin currents associated with the missing He-McKellar-Wilkens geometric quantum phase for $s=+1$ are
\begin{eqnarray}
\mathcal{I}=\frac{\varpi}{4\pi}+\sum_{\ell}\frac{\varpi}{4\pi}\frac{\left(\ell-\frac{\Phi_{\mathrm{MHMW}}}{2\pi}\right)}{\left|\ell-\frac{\Phi_{\mathrm{MHMW}}}{2\pi}\right|}.
\label{3.2}
\end{eqnarray}
For $s=-1$, we have
\begin{eqnarray}
\mathcal{I}=-\frac{\varpi}{4\pi}+\sum_{\ell}\frac{\varpi}{4\pi}\frac{\left(\ell+1-\frac{\Phi_{\mathrm{MHMW}}}{2\pi}\right)}{\left|\ell+1-\frac{\Phi_{\mathrm{MHMW}}}{2\pi}\right|}.
\label{3.3}
\end{eqnarray}

Hence, we can observe that the persistent spin currents (\ref{3.2}) and (\ref{3.3}) are influenced by the missing He-McKellar-Wilkens geometric quantum phase. Thereby, they are a periodic function of the geometric quantum phase $\Phi_{\mathrm{MHMW}}$.

\section{Aharonov-Bohm-type effect in a quantum ring}

In this section, we study the confinement of the neutral particle to an one-dimensional quantum ring of radius $R$ ($R\geq r_{0}$). According to Refs. \cite{ring,ring2,griff}, the Schr\"odinger-Pauli equation (\ref{1.1}) must be written in terms of a fixed radius $R$. In this way, we have  
\begin{eqnarray}
\mathcal{E}\psi&=&-\frac{1}{2m\,R^{2}}\frac{\partial^{2}\psi}{\partial\varphi^{2}}+\frac{i\,\sigma^{3}}{2m\,R^{2}}\frac{\partial\psi}{\partial\varphi}+\frac{1}{8m\,R^{2}}\,\psi-\frac{d\rho_{m}r_{0}^{2}\,\sigma^{3}}{2m\,R^{2}}\left(-i\frac{\partial}{\partial\varphi}-\frac{\sigma^{3}}{2}\right)\psi\nonumber\\
&+&\frac{d^{2}\rho_{m}^{2}r_{0}^{4}}{8m\,R^{2}}\,\psi+\frac{d\rho_{m}\,\sigma^{3}}{2m}\left(-i\frac{\partial}{\partial\varphi}-\frac{\sigma^{3}}{2}\right)\psi+\frac{d^{2}\rho_{m}^{2}R^{2}}{8m}\,\psi-\frac{d^{2}\rho_{m}^{2}r_{0}^{2}}{4m}\,\psi-\frac{d\rho_{m}}{2m}\,\psi.
\label{2.1}
\end{eqnarray}

Observe that the solution to Eq. (\ref{2.1}) can be given by
\begin{eqnarray}
\psi\left(\varphi\right)=\left(
\begin{array}{c}
f_{+}\left(\varphi\right)\\
f_{-}\left(\varphi\right)\\
\end{array}\right).
\label{2.2}
\end{eqnarray}
After substituting the wave function (\ref{2.2}) into Eq. (\ref{2.1}), we obtain two non-coupled equations for $f_{+}\left(r\right)$ and $f_{-}\left(r\right)$:
\begin{eqnarray}
\frac{d^{2}f_{s}}{d\varphi^{2}}-2i\,\nu\,\frac{df_{s}}{d\varphi}+\beta\,f_{s}=0,
\label{2.3}
\end{eqnarray}
where the equation for the function $f_{+}\left(r\right)$ is indicated by $s=+1$, while the equation for the function $f_{-}\left(r\right)$ is indicated by  $s=-1$. In addition, we have defined $\nu$ and $\beta$ as follows:
\begin{eqnarray}
\nu&=&\frac{\Phi_{\mathrm{MHMW}}}{2\pi}+\frac{s}{2}-\frac{s}{2}\,m\,\varpi\,R^{2};\nonumber\\
\beta&=&2mR^{2}\mathcal{E}-\nu^{2}+m\,\varpi\,R^{2},
\label{2.4}
\end{eqnarray}
where $\varpi$ has been defined in Eq. (\ref{1.6}). Thereby, the solution to Eq. (\ref{2.3}) is given by
\begin{eqnarray}
f_{s}\left(\varphi\right)=a_{0}\,e^{\eta\,\varphi},
\label{2.5}
\end{eqnarray}
where $a_{0}$ is a constant and $\eta$ is a parameter to be determined. By substituting (\ref{2.5}) into Eq. (\ref{2.3}), we find $\eta=i\left(\nu\pm\sqrt{\nu^{2}+\beta}\right)$, and then, the wave function (\ref{2.5}) becomes
\begin{eqnarray}
f_{s}\left(\varphi\right)=a_{0}\,e^{i\left(\nu\pm\sqrt{\nu^{2}+\beta}\right)\varphi}.
\label{2.6}
\end{eqnarray}

For an one-dimensional quantum ring, we have that $0\leq\varphi\leq2\pi$. Since the neutral particle is a spin-$1/2$ particle, hence, we have the boundary condition:
\begin{eqnarray}
f_{s}\left(\varphi+2\pi\right)=e^{i\,\pi}\,f_{s}\left(\varphi\right).
\label{2.7}
\end{eqnarray}
Next, by substituting the wave function (\ref{2.6}) into Eq. (\ref{2.7}), we find the relation:
\begin{eqnarray}
\nu\pm\sqrt{\nu^{2}+\beta}=\ell+\frac{1}{2},
\label{2.8}
\end{eqnarray}
where $\ell=0,\pm1,\pm2,\pm3,\ldots$, and thus, $j=\ell+\frac{1}{2}=\pm\frac{1}{2},\pm\frac{3}{2},\ldots$. Hence, after substituting the parameters $\nu$ and $\beta$ defined in Eq. (\ref{2.4}) into Eq. (\ref{2.8}), we obtain
\begin{eqnarray}
\mathcal{E}_{\ell}=\frac{1}{2m\,R^{2}}\left[\ell+\frac{1}{2}\left(1-s\right)-\frac{\Phi_{\mathrm{MHMW}}}{2\pi}+\frac{s}{2}\,m\,\varpi\,R^{2}\right]^{2}-\frac{\varpi}{2}.
\label{2.9}
\end{eqnarray}

Therefore, Eq. (\ref{2.9}) yields the energy levels of a neutral particle confined to a one-dimensional quantum ring. We can also observe that the energy levels (\ref{2.9}) are influenced by the missing He-McKellar-Wilkens geometric quantum phase $\Phi_{\mathrm{MHMW}}$, which also yields an Aharonov-Bohm-type effect \cite{pesk,griff,ring2}. Moreover, the spectrum of energy (\ref{2.9}) is a periodic function of the missing He-McKellar-Wilkens geometric quantum phase $\Phi_{\mathrm{MHMW}}$. For instance, with the periodicity $\phi_{0}=2\pi$ and $s=+1$, the quantum states determined by $\left\{\ell,\,s=+1\right\}$ and $\left\{\ell-1,\,s=+1\right\}$ are degenerate, i.e., $\mathcal{E}_{\ell}\left(\Phi_{\mathrm{MHMW}}+2\pi\right)=\mathcal{E}_{\ell-1}\left(\Phi_{\mathrm{MHMW}}\right)$. If we consider $s=-1$ and $\phi_{0}=2\pi$, we have the quantum states determined by $\left\{\ell,\,s=-1\right\}$ and $\left\{\ell-1,\,s=-1\right\}$ are also degenerated, where $\mathcal{E}_{\ell}\left(\Phi_{\mathrm{MHMW}}+2\pi\right)=\mathcal{E}_{\ell-1}\left(\Phi_{\mathrm{MHMW}}\right)$.

\subsection{Persistent spin currents}

Since the energy levels (\ref{2.9}) are a periodic function of the geometric quantum phase $\Phi_{\mathrm{MHMW}}$, we can also have the appearance of persistent spin currents in the one-dimensional quantum ring. In this case, by assuming that the temperature is $T=0$, the Byers-Yang relation \cite{by,dantas,bf3,bf18} becomes:
\begin{eqnarray}
\mathcal{I}=-\sum_{\ell}\frac{\partial \mathcal{E}_{\ell}}{\partial\Phi_{\mathrm{MHMW}}}.
\label{4.1}
\end{eqnarray}

Then, from the energy levels (\ref{2.9}), the persistent spin currents associated with the missing He-McKellar-Wilkens geometric quantum phase for $s=+1$ are
\begin{eqnarray}
\mathcal{I}=\frac{1}{2\pi\,m\,R^{2}}\,\sum_{\ell}\left[\ell-\frac{\Phi_{\mathrm{MHMW}}}{2\pi}+\frac{1}{2}\,m\,\varpi\,R^{2}\right],
\label{4.2}
\end{eqnarray}
while for $s=-1$:
\begin{eqnarray}
\mathcal{I}=\frac{1}{2\pi\,m\,R^{2}}\,\sum_{\ell}\left[\ell+1-\frac{\Phi_{\mathrm{MHMW}}}{2\pi}-\frac{1}{2}\,m\,\varpi\,R^{2}\right].
\label{4.3}
\end{eqnarray}

Therefore, the persistent spin currents (\ref{4.2}) and (\ref{4.3}) depend on the missing He-McKellar-Wilkens geometric quantum phase, which means that they are a periodic function of the geometric quantum phase $\Phi_{\mathrm{MHMW}}$.

\section{Conclusions}

By assuming the existence of magnetic charges, we have revisited the interaction of the permanent electric dipole moment of the neutral particle with the inhomogeneous magnetic field produced by a uniform distribution of the magnetic charges inside an infinitely long non-conducting cylinder, which possesses an inner radius $r_{0}$. We have assumed the existence of an infinity wall at $r=r_{0}$ \cite{b}. Then, we have shown a case where the Landau-He-McKellar-Wilkens levels \cite{lin} can be achieved even though an infinity wall exists at $r=r_{0}$. We have seen that the Landau-He-McKellar-Wilkens levels are modified by the missing He-McKellar-Wilkens geometric phase $\Phi_{\mathrm{MHMW}}$. Therefore, the missing magnetic charge per unit length inside the region $r\,<\,r_{0}$ influences the spectrum of energy, where the geometric quantum phase $\Phi_{\mathrm{MHMW}}$ breaks the degeneracy of the Landau-He-McKellar-Wilkens levels \cite{lin} and yields an Aharonov-Bohm-type effect for bound states \cite{ab,pesk,b}. Moreover, the energy levels are a periodic function of the missing He-McKellar-Wilkens geometric phase $\Phi_{\mathrm{MHMW}}$. From this periodic behaviour of the energy levels, we have calculated the persistent spin currents, and then, we have seen that the persistent spin currents are a periodic function of the geometric quantum phase $\Phi_{\mathrm{MHMW}}$.

We have also studied the confinement of the neutral particle to a one-dimensional quantum ring. We have seen that the energy levels depend on the missing He-McKellar-Wilkens geometric phase $\Phi_{\mathrm{MHMW}}$. This dependence of the missing He-McKellar-Wilkens geometric phase $\Phi_{\mathrm{MHMW}}$ has also given rise to an analogue of the Aharonov-Bohm effect for bound states \cite{ab,pesk,b}, and shown that the energy levels are a periodic function of the missing He-McKellar-Wilkens geometric phase $\Phi_{\mathrm{MHMW}}$. Further, we have calculated the persistent spin currents that appear in the one-dimensional quantum ring, and shown that the persistent spin currents are also a periodic function of the geometric quantum phase $\Phi_{\mathrm{MHMW}}$

\acknowledgments{The author would like to thank CNPq for financial support.}

\section*{Data accessibility statement}

This work does not have any experimental data.

\section*{Competing interests}

We have no competing interests.

\end{document}